\documentclass[]{elsarticle}

\journal{---xxxxxx---}
\usepackage[utf8]{inputenc}
\usepackage{amsfonts}

\begin{document}

\begin{frontmatter}

\title{An efficient method for tracing high-resolution invariant manifolds of three-dimensional flows}
\author[USP]{D. Ciro\corref{corr-auth}}
\cortext[corr-auth]{davidcirotaborda@usp.br}
\author[GA]{T. E. Evans}
\author[USP]{I. L. Caldas}

\address[USP]{Universidade de São Paulo, 05508-090, São Paulo, Brazil.}
\address[GA]{General Atomics, PO Box 85608, San Diego, CA 92186-5608, USA}

\begin{abstract}
In Hamiltonian systems subjected to periodic perturbations the stable and unstable manifolds of the unstable periodic orbits provide the dynamical ``skeleton'' that drives the mixing process and bounds the chaotic regions of the phase space. Determining the behavior of these objects is valuable in physical applications involving asymmetric solenoidal fields or time-dependent Hamiltonian systems. Here we introduce a simple method to calculate an unstable periodic orbit given an initial guess on its position. Then we present an efficient adaptive method to build its high-resolution invariant manifolds to arbitrary length and compare it to a random sampling method with the same computational cost. The adaptive method gives a high-quality representation of the manifolds and reveals fine details that become lost in the random sampling method. Finally, we introduce an approximation to the adaptive method to build the manifolds avoiding redundant calculations and reducing logarithmically the number of computations needed to represent these surfaces.
\end{abstract}

\end{frontmatter}

%\linenumbers

\section{Introduction}
In time-independent Hamiltonian systems the value of the Hamiltonian function is conserved by the solutions of the equations of motion~\cite{goldstein80}. This conservation guarantees that the orbits of the system are attached to suitable level-sets of the Hamiltonian. In the near-integrable case, where small time-dependent perturbations are applied, this condition is relaxed and some orbits are allowed to wander inside a finite region with a complicated aperiodic motion known as Hamiltonian chaos~\cite{zaslavsky07}.

Much of the complicated behavior on the chaotic region is caused by the proliferation of unstable periodic orbits due to the horseshoe mechanism~\cite{guckenheimer83} caused by the intersection of the stable and unstable manifolds of one or various saddle orbits~\cite{portela07}.

In physical situations, the identification of unstable periodic orbits and the calculation of invariant manifolds are important to determine relevant geometric features of a steady vector field. In general, the differential equations defining the tracers of a solenoidal field can be expressed in Hamiltonian form, with the time being a suitable spatial coordinate~\cite{biskamp92}. The magnetic field ($\nabla\cdot\vec B=0$) and the incompressible fluid velocity ($\nabla\cdot\vec V=0$) belong to this category. Then, in any asymmetric situation we can expect the emergence of chaotic regions about some saddle orbit or a resonant surface~\cite{zaslavsky07}.

To determine these structures in a particular situation it is important to develop efficient numerical methods to build the stable and unstable manifolds of a given saddle orbit. In some situations, the geometry of the manifolds can be estimated through the Poincaré plot of a large collection of orbits close to the saddle. However, without an ordering scheme, this method is limited in resolution and does not allow us to improve the representation of the manifold in a controlled fashion.

In the present work we introduce a simple method to calculate the saddle orbit for a near-integrable periodic Hamiltonian and an efficient adaptive method to calculate the corresponding manifolds to arbitrary length and resolution, reducing superfluous calculations and exploiting the fundamental properties of the Poincar\'e map. We also present an efficient approximation method that reduces the computational cost of the adaptive algorithm in situations where the manifolds must be spanned by a large number of orbits or must be followed for many periods. The use of the methods is illustrated though a Hamiltonian system that represents the same topological features as a single-null magnetic configuration common in magnetic confinement devices~\cite{kadomtsev92}.

\section{The perturbed saddle}
Hamiltonian dynamical systems with $1+1/2$ degrees of freedom can be written in the form
\begin{equation}\label{hamilton_eqs}
 \frac{dx}{dt} = \frac{\partial H}{\partial y}\mbox{ , }\frac{dy}{dt} = -\frac{\partial H}{\partial x},
\end{equation}
where $H(x,y,t)$ is the Hamiltonian function. In the time-independent case $H(x,y)$ is a constant of the motion and the fixed points of the system can be obtained by requiring a vanishing flow $dx/dt = dy/dt = 0$ and solving (\ref{hamilton_eqs}) for $x$ and $y$ simultaneously. The number of fixed points depends on the form of $H(x,y)$. These points can be classified in \emph{saddles} and \emph{centers}, depending on the eigenvalues of the Jacobian matrix of the vector field\S. When a fixed point is a saddle the Jacobian matrix has real eigenvalues and there is an attracting and a repelling direction driving the orbits on its neighborhood. When a fixed point is a center the Jacobian has imaginary eigenvalues and the nearby orbits encircle the point.

If the Hamiltonian is a periodic function, for instance  $H(x,y,t)=H_0(x,y)+H_1(x,y,t)$, with $H_1(x,y,t+T)=H_1(x,y,t)$, its value is no longer a constant of the motion and the points where the field vanishes are no longer fixed points. Intuitively, the point $p_0(t)$, where $(\dot x,\dot y)$ vanishes at $t$ is slightly different from the point where it vanishes at $t+dt$, and these points are not, in general, contained in the same orbit.

In the more general time-dependent situation, the saddle corresponds to an unstable periodic orbit with the same period of the Hamiltonian function, but before introducing the method to calculate this curve, let us introduce some basic definitions that will ease the understanding of the following sections.

The \emph{Poincaré map} $M:U \rightarrow U$, is a function that takes any initial condition $p\in U$ in a given time $t$ and returns the point $M(p)\in U$ in the time $t+T$ obtained by solving the Hamilton equations. Likewise the inverse Poincaré map $M^{-1}$ returns the point $M^{-1}(P)\in U$ obtained by solving the equations for $t-T$. Clearly $M^{-1}\circ M = M\circ M^{-1} = \mathbb{I}$ is the identity map.

In the transversal section $U$, an \emph{orbit} $\gamma_p$, is the collection of the points obtained by applying the Poincaré map and its inverse to the point $p$.
$$
\gamma(p) = \{ ...,p_{-2},p_{-1},p_0,p_1,p_2,... \},
$$
where $p_{\pm n} = M^{\pm n}(p_0)$ and $n\in \mathbb{Z}^+$. Additionally, the sequences $\gamma^+(p_0)=\{p_0,p_1,p_2,...\}$ and $\gamma^-(p_0)=\{ ...,p_{-2},p_{-1},p_0\}$ are the \emph{forward} and \emph{backward orbit} from $p_0$. Clearly, the orbit $\gamma(p)$ can be obtained by applying repeatedly $M$ and $M^{-1}$ to any of its elements; then, the label $p$ can be any element $p_i$ of the list.

\subsection{Fixed points of $M$ (The three-mapping method)}\label{sec_saddle}
Clearly, a fixed point $p^*$ of the Poincaré map $M$ satisfies
\begin{equation}\label{fixed_point}
 M(p^*)=p^*.
\end{equation}
This means that this point is linked to a periodic orbit of the Hamiltonian system. However, in most situations, we do not have access to the analytical form of $M$ but only to its effects when applied to a point; then, to solve (\ref{fixed_point}) we need a method that only requires this information. Let us assume that we know a point $p_0$, that is close to the fixed point $p^*$. This is particularly true when the Hamiltonian is separable and the time-dependent part is small; then we can take $p_0$ to be the unperturbed saddle, which can be obtained analytically or numerically.

\begin{figure}[h]
 \centering
 \includegraphics[]{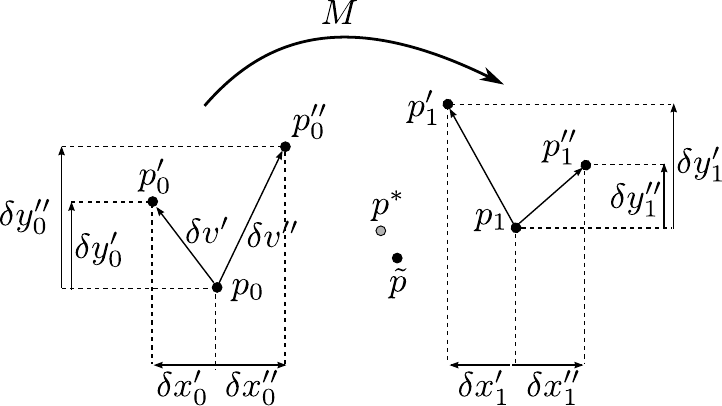}
 \caption{\label{f_find_fixed} Three points are mapped to approximate the Jacobian matrix and the fixed point.}
\end{figure}

By applying the map to the point $p_0$ we obtain $p_1$, that can be approximated by

\begin{equation}\label{eq_p1_approx}
 p_1 = (I-DM_{p^*})p^* + DM_{p^*}p_0 + O(\delta p^2).
 \end{equation}
Here, $p_0,p_1$ and $p^*$ are position vectors of the form $(x,y)^T$, $I$ is the $2\times 2$ identity matrix, $DT_{p^*}$ is the unknown Jacobian of $M$ evaluated in $p^*$ and $\delta p = p_0 - p^*$ is a small vector separating the saddle $p^*$ from $p_0$. In equation (\ref{eq_p1_approx}) we know only the coordinates of $p_0$ and $p_1$, so, we have \emph{six} unknowns, namely, the four entries of $DM_{p^*}$ and the coordinates of $p^*$. To obtain an approximated solution to the unknowns we need to apply the map to another two points $p_0' = p_0 + \delta v'$ and $p_0'' = p_0 + \delta v''$ close to $p_0$ (Fig.~\ref{f_find_fixed}), where $\delta v'$ and $\delta v''$ are small arbitrary vectors.
It is not difficult to show that

\begin{equation}\label{eq_jacobian}
 D_M = \left[
 \begin{array}{c c}
  \delta x_1' & \delta x_1''\\
  \delta y_1' & \delta y_1''
 \end{array}
 \right]
 \left[
 \begin{array}{c c}
  \delta x_0' & \delta x_0''\\
  \delta y_0' & \delta y_0''
 \end{array}
 \right]^{-1},
\end{equation}
is a second order approximation of the Jacobian matrix $DM_{p^*}$. Then we can use this to find an approximate solution to the fixed point
\begin{equation}\label{eq_fixed}
 \tilde p = (I-D_M)^{-1}(p_1 - D_Mp_0).
\end{equation}
If the first approximation $\tilde p$ does not satisfy $T(\tilde p)=\tilde p$, we can use it as a new initial value $p_0 = \tilde p$. Also we can choose $p_0'$ and $p_0''$, to be aligned with the eigenvectors of $D_M$. Applying (\ref{eq_jacobian}) we can get a new approximation for the Jacobian matrix and then (\ref{eq_fixed}) for a new approximation of the fixed point.

Upon repetition of this process we can get the fixed point of $M$ that is closest to the initial point $p_0$. Usually this requires less than five iterations. To illustrate this method we calculate the perturbed magnetic saddle of the dynamical system obtained from the following Hamiltonian function
\begin{equation}\label{eq_hamilton}
 H(x,y,t) = \frac{1}{4}(x^2 - 1)^2 + \frac{1}{2}y^2 +\epsilon\cos(x - \omega t).
\end{equation}
In the unperturbed case $\epsilon = 0$ this system has a saddle point at $(x^*,y^*) = (0,0)$ that we define to be $p_0$. The initial points $p_0'$ and $p_0''$ are chosen at a distance $10^{-7}$ from $p_0$ in a random direction.
After $5$ iterations of the three-point method we get the fixed point $p^* = \{-4.9067732(3)\times 10^{-9}, 0.031005460055075(9)\}$ and the Jacobian eigenvalues $\lambda_1=2.7182782448(1),\lambda_2=0.367879926(1)$. The digit in parenthesis indicate the decimal place where the fluctuations occur if the method continues further.

\begin{figure}[h]
 \centering
 \includegraphics[width=0.6\textwidth]{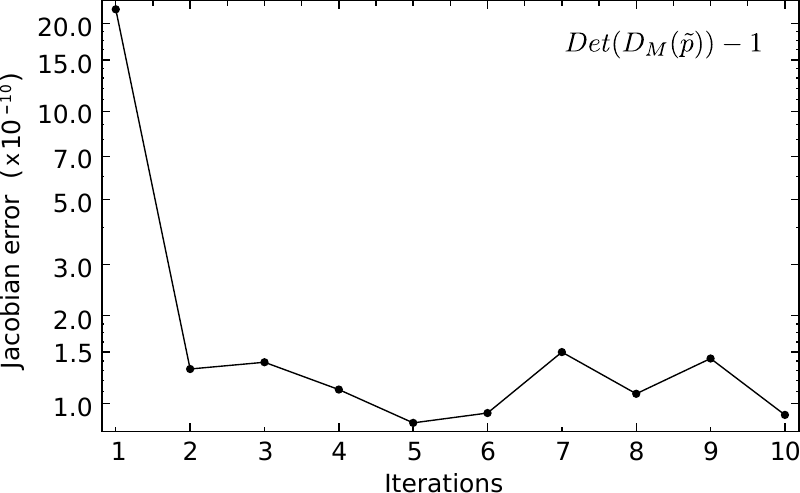}
 \caption{\label{f_determinant} The approximated Jacobian becomes a symplectic matrix ($Det(D_M)=1$) to a precision of $10$ digits in just two iterations of the three-three mapping method.}
\end{figure}

The Poincaré map of a Hamiltonian system must be area preserving, which means that the determinant of the Jacobian matrix, or the product of $\lambda_1$ and $\lambda_2$ must be unity. In this particular case we get $\lambda_1\lambda_2-1 = 0.9\times 10^{-11}$. In Fig.~\ref{f_determinant}, it is clear that the approximated Jacobian converges very rapidly to an almost symplectic matrix, with fluctuations bounded at $10^{-10}$.

\begin{figure}[h]
 \centering
 \includegraphics[]{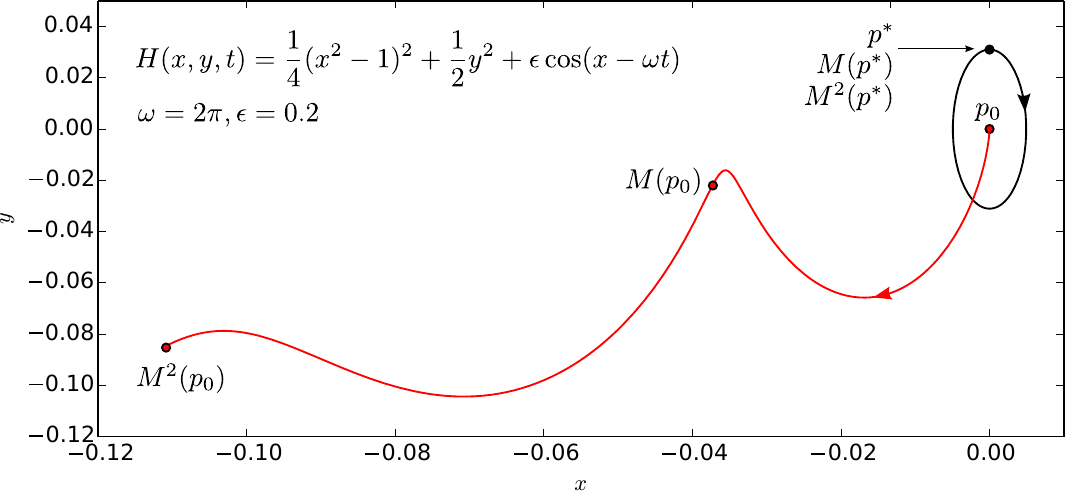}
 \caption{\label{f_saddle_orbits}Orbits starting from the point where the flow vanishes $p_0$, and from the point obtained by the three-mapping method $p^*$. The points over the orbit show the position after one period of the Hamiltonian function. The orbit from $p^*$ is periodic and $p^*$ is a fixed point of the Poincaré map $M$.}
\end{figure}

To illustrate that the obtained fixed point is linked to an unstable periodic orbit we can calculate the trajectories for the dynamical system. In Fig.~\ref{f_saddle_orbits} we see two orbits starting at two near initial conditions in $t=0$. The first point $p_0 = \{0,0\}$ is the point where $\dot x = \dot y = 0$ for $t=0$ at any value of $\epsilon$ and the point $p^*$ is obtained for $5$ iterations of the three-mapping method. Clearly $p_0$ is not a fixed point of the Poincaré map because its orbit is not periodic. From the fixed point $p^*$, the trajectory becomes a closed orbit with the period of the Hamiltonian function. This orbit correspond to the perturbed saddle, and nearby orbits will move away from it as they evolve.

The methods presented in this section can be easily modified to find higher order periodic orbits that are fixed points of the composite maps $M^n$. Also they are not restricted to area-preserving maps, and can be used for dissipative systems as well.

\section{Calculating the manifolds}
Once we know the perturbed saddle $X$, we can define its \emph{stable} and \emph{unstable manifolds}~\cite{guckenheimer83}
$$
 \mathcal{S}(X) = \{p \in U | M^n(p)\rightarrow X \mbox{ as } n\rightarrow \infty \}.
$$
$$
 \mathcal{U}(X) = \{p \in U | M^{-n}(p)\rightarrow X \mbox{ as } n\rightarrow \infty \}.
$$
In other words $\mathcal{S}(X)$ is the set of all the points pertaining to orbits evolving towards the X-point, and $\mathcal{U}(X)$ is the collection of the points pertaining to orbits coming from the X-point. Clearly, the saddle does not belong to any of these orbits but it belongs to both manifolds.

These manifolds can be depicted by calculating the orbits from a large number of starting points filling randomly a small region around the saddle point $X$ (random sampling) or a large region covering the chaotic region ~\cite{da-silva02} (sprinkler method). The initial conditions can also be uniformly aligned with the saddle eigenvectors in a very small neighborhood (uniform sampling). In neighborhood of the saddle-point orbits are governed by the Jacobian of the vector field, and the manifolds are aligned with its eigenvectors. To obtain the unstable manifold the Poincaré map $M$ is iterated over the initial conditions in the direction with eigenvalue larger than one. For the stable one, the inverse map $M^{-1}$ is iterated over the initial conditions of the direction with eigenvalue less than one.

\begin{figure}[h]
 \centering
 \includegraphics[]{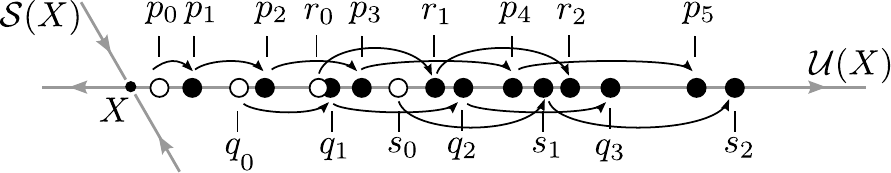}
 \caption{\label{f_uni_grid} The unstable manifold is traced by the orbits departing from $X$. The open circles are uniformly spaced initial conditions.}
\end{figure}

Each point of these orbits is very close or belongs to the manifold and a sufficiently large number of orbits will allow us to trace the underlying manifold (Fig.~\ref{f_uni_grid}).

The random and uniform sampling techniques, although simple, have some drawbacks. The length of the segment of initial conditions grows exponentially fast in time and the critical features of the manifolds appear when the segment gets stretched and folded repeatedly. In other words, the number of initial conditions must be sufficiently large as to maintain a continuous appearance after a given number of periods. Additionally, if the segment is sufficiently close to the saddle and its length is not chosen appropriately, the points closer to the saddle will be mapped back inside the segment, disordering the initial configuration.
Finally, some regions of the segment will suffer more stretching than others; then, an initially uniform distribution of points will grow unevenly, leading to sparse regions and regions with superfluous points. Clearly, before calculating the manifolds there is no knowledge about the regions that suffer from more stretching and the only way to maintain a continuous appearance with a random sampling or uniform sampling is to increase the number of initial conditions.

Finally, this method does not lead to an organized set of points that can be joined smoothly with a single line that represents the continuous manifold; consequently, we may loose track of the manifold in the most critical regions if the points get too scattered.

\section{The "exact" adaptive approach}
Consider that we want to represent $\mathcal{U}(X)$ and $\mathcal{S}(X)$ by sequences of ordered points
\begin{eqnarray}
 && \bar\mathcal{U} = \{ u_1,u_2,u_3... \}, \\
 && \bar\mathcal{S} = \{ s_1,s_2,s_3... \}.
\end{eqnarray}
A good method to trace the manifolds should return a reasonably uniform distribution of ordered points along the manifold. The ordering is important because in most cases each manifold is sufficiently complicated to become very close to itself (without crossing), so we need to know whether two close points are neighbors or not. In an ordered case we can join the points with a smooth line that mimics the manifold in contrast to the random or uniform sampling methods where the points are disordered and can not be joined.

\subsection{Ordering the manifold}
Each point in the manifold is an element of an orbit. We can identify each point in $\bar\mathcal{U}$ by a given tag $t_i$ corresponding to its orbit. This give us a list of tags for the manifold
\begin{equation}
 \mathcal{TU} = \{t_1,t_2,t_3,...\}.
\end{equation}
If $\bar\mathcal{U}$ is composed by $n$ different orbits, the tags in $\mathcal{TU}$ will follow a periodic sequence of $n$ different tags. This happens because the map $M$ preserves the ordering of the orbits along the manifolds. Then we only need the periodic sequence $\{t_1,t_2,...,t_n\}$ to order the manifold.

We can exploit this fact to preserve the orbits separated, and when required, build the manifold using the periodic sequence in $\mathcal{TU}$.

\begin{figure}[h]
 \centering
 \includegraphics[]{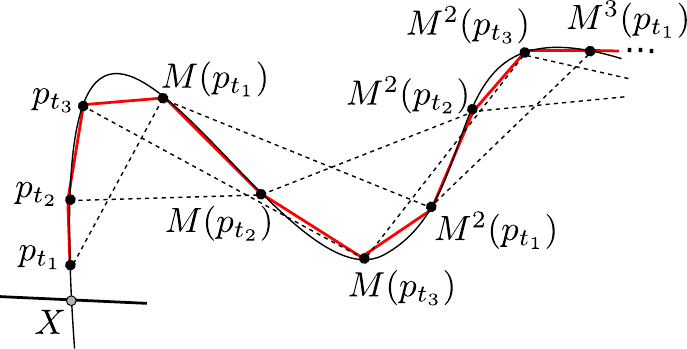}
 \caption{\label{f_approx_manif} The corresponding points of three organized orbits can be joined (red) to mimic the underlying manifold (black). The dashed lines follow the individual orbits.}
\end{figure}

Let us define a list of orbits
\begin{equation}
 \mathcal{A}_n = [\gamma_1,\gamma_2,...,\gamma_m],
\end{equation}
where each $\gamma_i$ is a forward orbit starting on the unstable manifold close to $X$,
\begin{equation}
 \gamma_i = [p_i,M(p_i),...,M^N(p_i)]^T.
\end{equation}
The orbits in $\mathcal{A}_n$ do not need to be organized in any logical way, but this information can be found in the ordering tags $\mathcal{T}_n$
\begin{equation}
 \mathcal{T}_n = \{t_1,t_2,...,t_n\},
\end{equation}
such that $t_i\neq t_j$ if $i\neq j$ and $t_i\in \{1,2,...,n\}$. The list $\mathcal{T}_n$ is a permutation of the numbers $\{1,2,...,n\}$ that defines the right ordering of the orbits in $\mathcal{A}_n$ such that we can join the points to follow the manifold in a consistent form (Fig.~\ref{f_approx_manif}).

Clearly, the $n$'th refinement of the unstable manifold can be written as
\begin{equation}
 U_n = \{p_{t_1},p_{t_2},...,p_{t_n},M(p_{t_1}),M(p_{t_2}),...,M^m(p_{t_n})\}.
\end{equation}

\subsection{Algorithm to build the manifold}
To build the manifold we require an adaptive process that introduces new orbits containing points that improve the manifold tracing in a given sparse region. For this, we define the maximum separation between consecutive points $d_{max}$, and require the points in the manifold to be separated by \emph{at most} that distance.

\subsubsection{Initialization}
Assume that we have a starting orbit $\gamma_0 = \{p,M(p),...,M^{m+1}(p)\}$, such that $p$ and $M(p)$ are very close to the magnetic saddle and it is safe to say that $U$ is a straight line between them. Then, we define the orbits
$$
\gamma_1 = \{p,M(p),...,M^{m}(p)\}, \gamma_2 = \{M(p),...,M^{m+1}(p)\},
$$
to initialize the list of orbits as
$$
\mathcal{A}_2 = \{\gamma_1,\gamma_2\},
$$
and the ordering tags as
$$
\mathcal{T}_2 = \{1,2\}.
$$

\subsubsection{Refinement procedure}
The list $\mathcal{A}_n$ contains $n$ different orbits, each with $m$ points. The elements $a_{i,j} = \gamma_{i,j}$ are the $j$'th element of the orbit $i$. The ordering list $\mathcal{T}_n$ is a list with $n$ different tags and $N$ is the number of new orbits to be calculated.

The refinement of the manifold consists in introducing orbits between two neighbor orbits that become too widely separated after a given number of cycles. The initial condition of the new orbit is at the midpoint between the initial conditions of the neighbor orbits. The new tag $n+1$ must be introduced at the corresponding position in the tag list $\mathcal{T}_n$.

Consider the following pseudo-code for this procedure:

\begin{verbatim}
k0 = n+1;               /// initial tag
kf = n+N;               /// final tag
For(k=k0, k<=kf, k++){ 
  i = 1; j = 1; 
                        /// search for a large gap
  While(Norm(A[T[i+1]][j]-A[T[i]][j]) < d_max){
    If(i<n){ i++ };     /// move in the list of orbits 
    Else{ i = 1; j++ }; /// move inside the orbits 
  };
                        /// A[i] and A[i+1] become separated
                            after j cycles
                        /// calc. new initial condition.
  p = 0.5*(A[T[i+1]][1]+A[T[i]][1]);
  orbit = Orbit(p,m);   /// calculate the new orbit from p 
  Append(A,orbit);      /// append new orbit to A 
  Insert(T,k,i+1);      /// insert tag "k" in position i+1 of T.
  n++;                  /// increase the dimension of A
 }
\end{verbatim}
 
Where \verb T[n]  is the list of ordering tags and \verb A[n][m]  is the array of points defining the manifold. The function \verb Orbit(point[2],length)  returns an orbit of length \verb length  from the initial point \verb point  .\\The function \verb Append(Array[n][m],orbit[m])  returns an array \verb A[n+1][m]  such that \verb A[n+1][i]=orbit[i]  , and, finally, the function \verb Insert(T[n],k,i+1)  returns the list \verb T[n+1]  with \verb k  at the index \verb i+1  and the next elements displaced one index forward with respect to their original position.

The use of an ordering list $\mathcal{T}_n$, avoids rewriting the whole manifold $\mathcal{A}$ each time that a new orbit is calculated. Instead we only need to rewrite the list of tags, a computationally less expensive job.
 
\section{Example application}

\begin{figure}[h]
 \centering
 \includegraphics[]{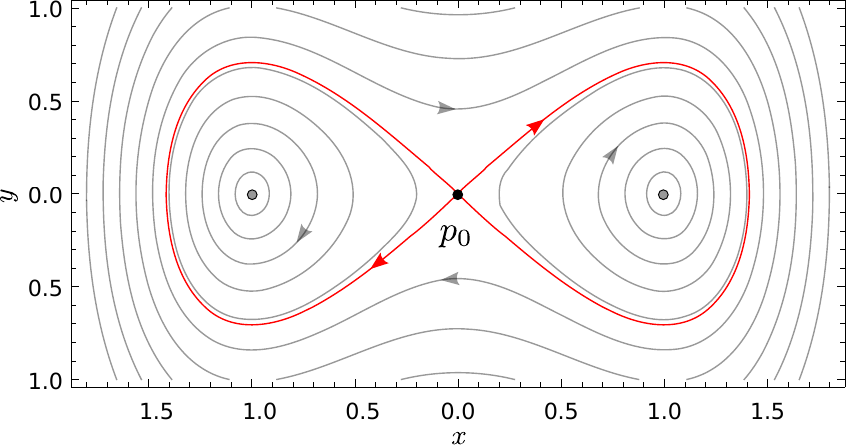}
 \caption{\label{f_contours}Level curves of the Hamiltonian function (\ref{eq_hamiltonian}) for $\epsilon = 0$. The arrows indicate the direction of the field (\ref{eq_x},\ref{eq_y})}
\end{figure}

Consider again the Hamiltonian
\begin{equation}\label{eq_hamiltonian}
 H(x,y,t) = \frac{1}{4}(x^2-1)^2 + \frac{1}{2}y^2 + \epsilon\cos(x - \omega t),
\end{equation}
leading to non-autonomous dynamical system
\begin{eqnarray}
 \dot x &=& y, \label{eq_x}\\ 
 \dot y &=& x(1-x^2) + \epsilon\sin(x - \omega t) \label{eq_y}.
\end{eqnarray}
For $\epsilon = 0$ the Hamiltonian function is a constant of the motion and its level curves correspond to the parametric curves of the solutions to the system (\ref{eq_x}, \ref{eq_y})(Fig.~\ref{f_contours}). This system possesses a single saddle point between two centers and a separatrix defined by two homoclinic orbits.

\begin{figure}[h]
 \centering
 \includegraphics[width=\textwidth]{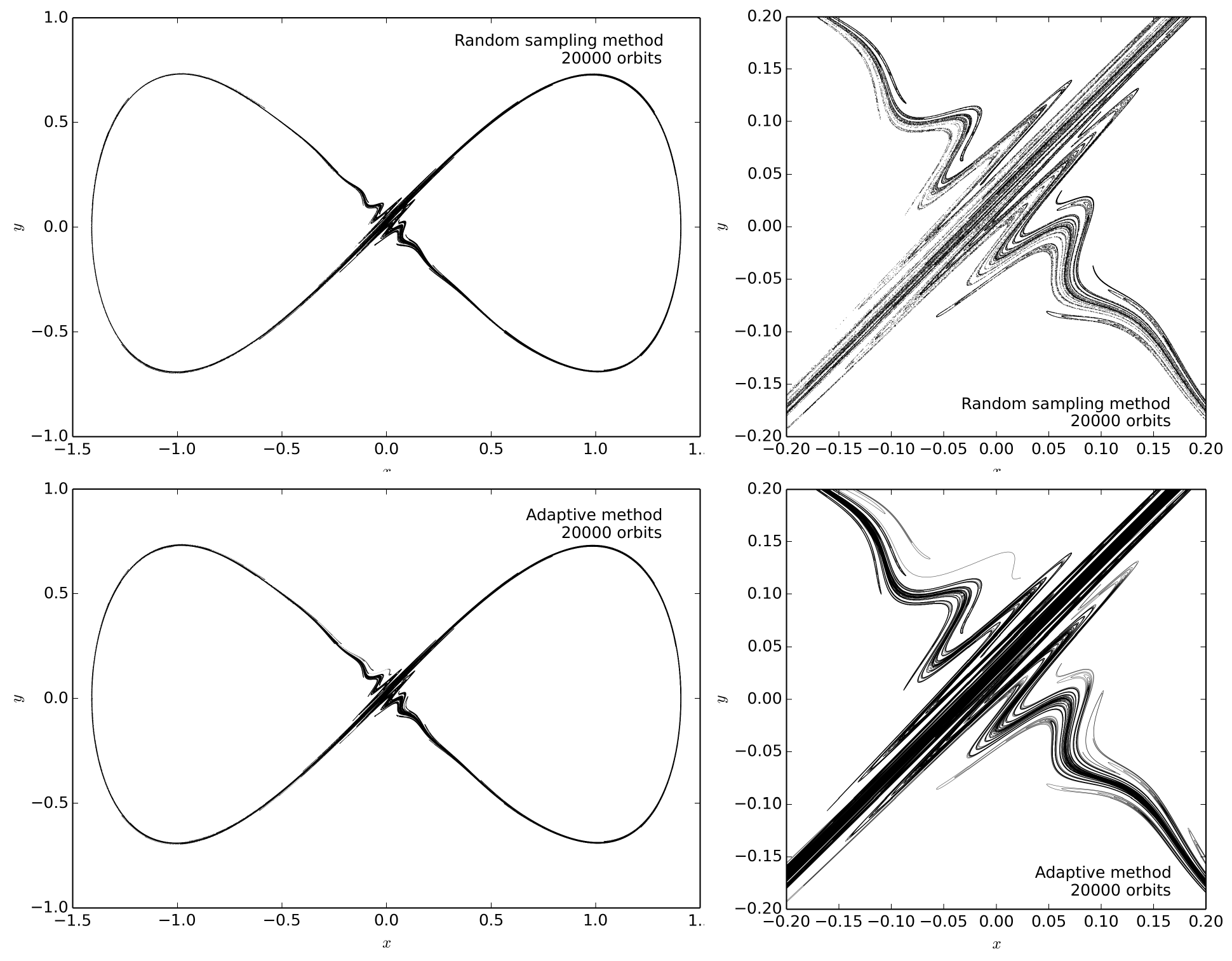}
 \caption{\label{comparison1}Unstable manifold calculated by the random sampling method (top) and the adaptive method (bottom). In both cases the manifold contains $8\times 10^5$ points but in the adaptive method they are organized and can be joined with a line.}
\end{figure}

As mentioned in Section~\ref{sec_saddle}, for $\epsilon\neq 0$ the saddle point continues to exist, but is no longer at the origin, even if $\dot x = \dot y =0$ at that point. For $\epsilon = 0.2$ this point is slightly displaced to 
$$p^*=\{4.9067732(3)\times 10^{-9},0.031005460055075(9)\}$$
and the eigenvectors of $DM_p^*$ are
$$v_1 = (0.703856076086467, 0.710342610404417)^T,$$
$$v_2=(-0.7038551192759112, 0.7103435584765255)^T.$$
To calculate the unstable manifold with the traditional method we build a small square of initial conditions of size $10^{-5}$ about $p^*$. Then we iterate the Poincaré map for each one to build orbits with $m = 40$ points (Fig.~\ref{comparison1}).

\begin{figure}[h]
 \centering
 \includegraphics[width=\textwidth]{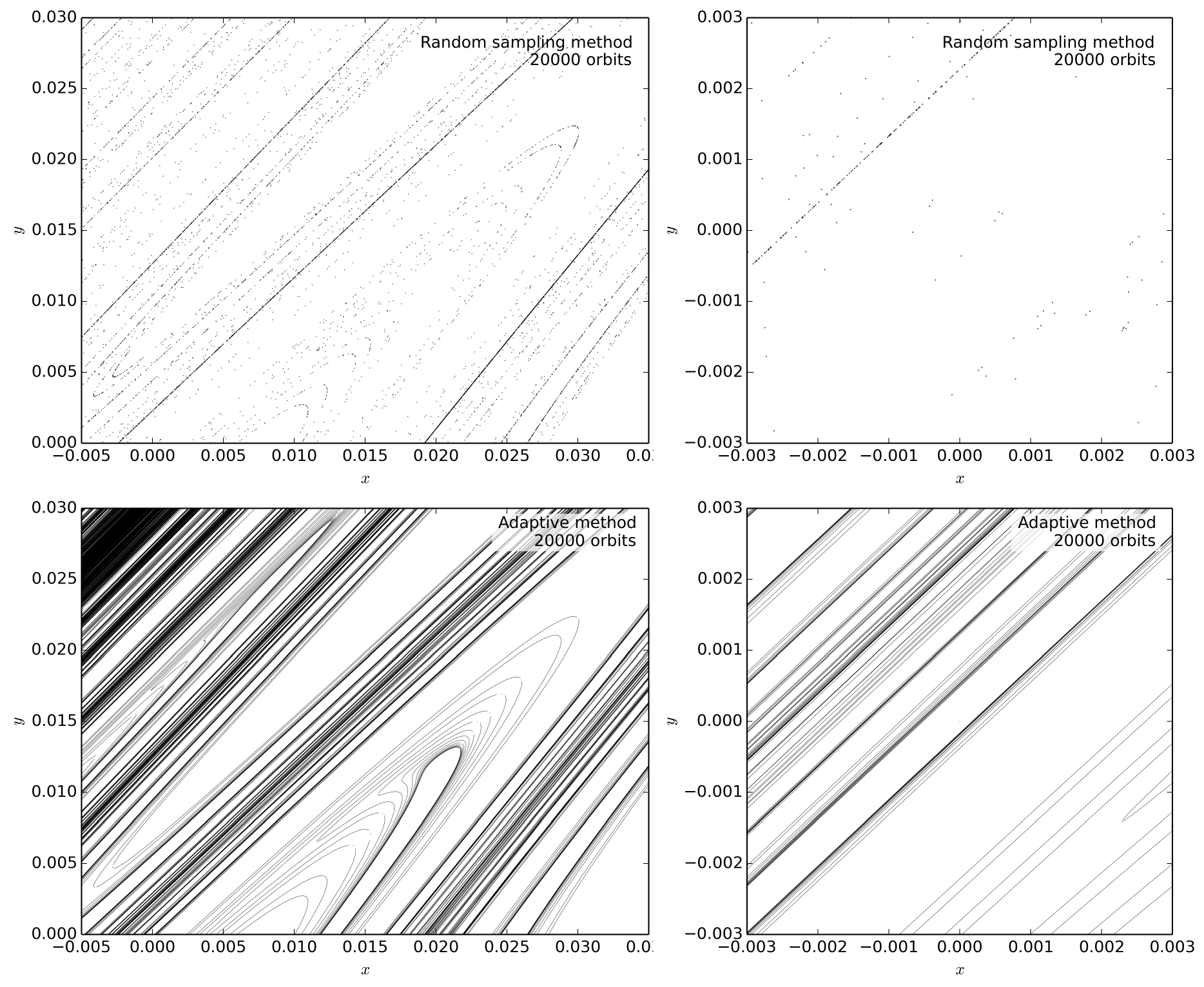}
 \caption{\label{comparison2}Details of the unstable manifold from the random sampling (top) and the adaptive method (bottom). At this scale the even distribution of points from the adaptive method reveals fine details of the manifold that develop in a sparse region of the random sampling.}
\end{figure}

In Fig.~\ref{comparison1} we show the unstable manifold using $2\times10^4$ orbits, which give us a total of $8\times10^5$ points. This is done using the random sampling and the adaptive approach. As discussed before, the adaptive method provides ordered points to describe the manifold and we can join them with lines to understand better its behavior. In Fig.~\ref{comparison1} the ordering does not appear to provide additional information about the manifold, but as we zoom into a smaller region the advantage of the adaptive approach becomes evident (Fig.~\ref{comparison2}).

The level of detail of the adaptive method in Fig.~\ref{comparison2} is due to the construction mechanism. Each time we find a large gap in the manifold a new orbit is created from an initial condition that will lead to a point in the desired location. This prevents the accumulation of points in well resolved regions and successively fills the sparse regions providing a good distribution of points in regions where other methods rely on the number of initial conditions near the saddle.

\section{The "approximated" adaptive approach}
The "exact" adaptive approach returns an ordered set of well distributed points along the manifold. However, every component orbit is calculated from a neighborhood of the saddle point, where the dynamics is very slow. Consequently, there is a considerable amount of computation required to leave this region and reach the region where the condition $|u_{i+1}-u_i|<d_{max}$ is violated.

As we increase the length and complexity of the manifold the number of points required to describe it may become very large and the amount of computation becomes an important issue. Thus, it is desirable to avoid the calculation of unnecessary cycles. This involves creating initial conditions far from the saddle point, in regions that can not described by a linear map.

\begin{figure}[h]
 \centering
 \includegraphics[]{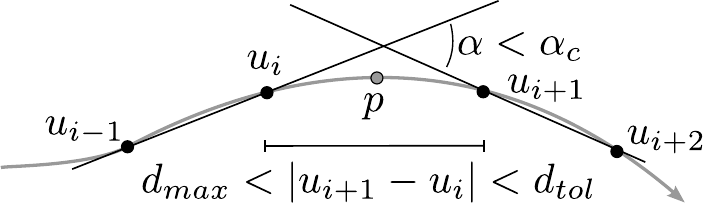}
 \caption{\label{f_new_init} The new initial condition $p$ is calculated between points not too distant or close and in a sufficiently smooth region of the manifold.}
\end{figure}

Consider that we want to calculate $p$ between $u_i$ and $u_{i+1}$, such that $p$ is sufficiently close to the manifold $\mathcal{U}(X)$ and consequently the rest of the forward orbit $\{M^n(p)\}$. However, we do not want to calculate the points of the orbit from the saddle neighborhood to $p$. To do this, we assume that the manifold is a smooth parametric curve $\Gamma(s)=(x_u(s),y_u(s))$ passing through the four points $\{u_{i-1},u_i,u_{i+1},u_{i+2}\}$ at the parameter values $s=\{0,1,2,3\}$ respectively. If we choose this curve to be cubic in $s$ it is not difficult to show that it takes the form
\begin{eqnarray}
 \Gamma(s) &=& (-u_{i-1} + 3 u_i - 3 u_{i+1} + u_{i+2})\frac{s^3}{6} + 
   (2 u_{i-1} - 5 u_i + 4 u_{i+1} - u_{i+2})\frac{s^2}{2}\nonumber\\
   && + (-11 u_{i-1} + 18 u_i - 9 u_{i+1} + 2 u_{i+2})\frac{s}{6} + u_{i-1}.
\end{eqnarray}
We can evaluate $\Gamma(s)$ in any place between $t=1$ and $t=2$. Typically we can simply choose $t=1.5$, which leads to
\begin{equation}
 p = \Gamma(1.5) = \frac{1}{16} (-u_{i-1} + 9 u_i + 9 u_{i+1} - u_{i+2}).
\end{equation}
Even when the parameter values $t=\{1,2,3,4\}$ were introduced somewhat arbitrarily, when the manifold is sufficiently smooth between between $u_{i-1}$ and $u_{i+2}$ it is well approximated even if we do not know the correct parametrization. To ensure this, the intermediate points are only calculated when the angle between the vectors $u_i-u_{i-1}$ and $u_{i+2}-u_{i+1}$ is smaller than some predefined value $\alpha_c$. Also, the distance between $u_i$ and $u_{i+1}$ must be larger than the required maximum separation $d_{max}$ and smaller than a tolerance distance $d_{tol}$ (Fig.~\ref{f_new_init}). Similar conditions are used in~\cite{hobson92} to keep track of the spatial resolution when growing the manifolds with primary segments.

With these restrictions we guarantee that the cubic approximation gives a reliable point in a region that is not too dense or too sparse. Using the cubic approximation in very sparse regions may lead to small spurious oscillations in the manifold caused by the inclusion of nearby orbits that do not converge to the saddle point.

It can be shown that the number of Poincaré cycles required by the approximated method is about
\begin{equation}
 N_c \approx N\left[m - \lambda Ln\left(r N\right) \right],
\end{equation}
where $m$ is the number of points in each orbit of the ``exact'' adaptive method, and $N$ is the number of orbits. Here, $\lambda$ characterizes the elongation of any small segment of initial conditions near the saddle and $r$ characterizes their distribution in the stripe.

\begin{figure}[h]
 \centering
 \includegraphics[width=\textwidth]{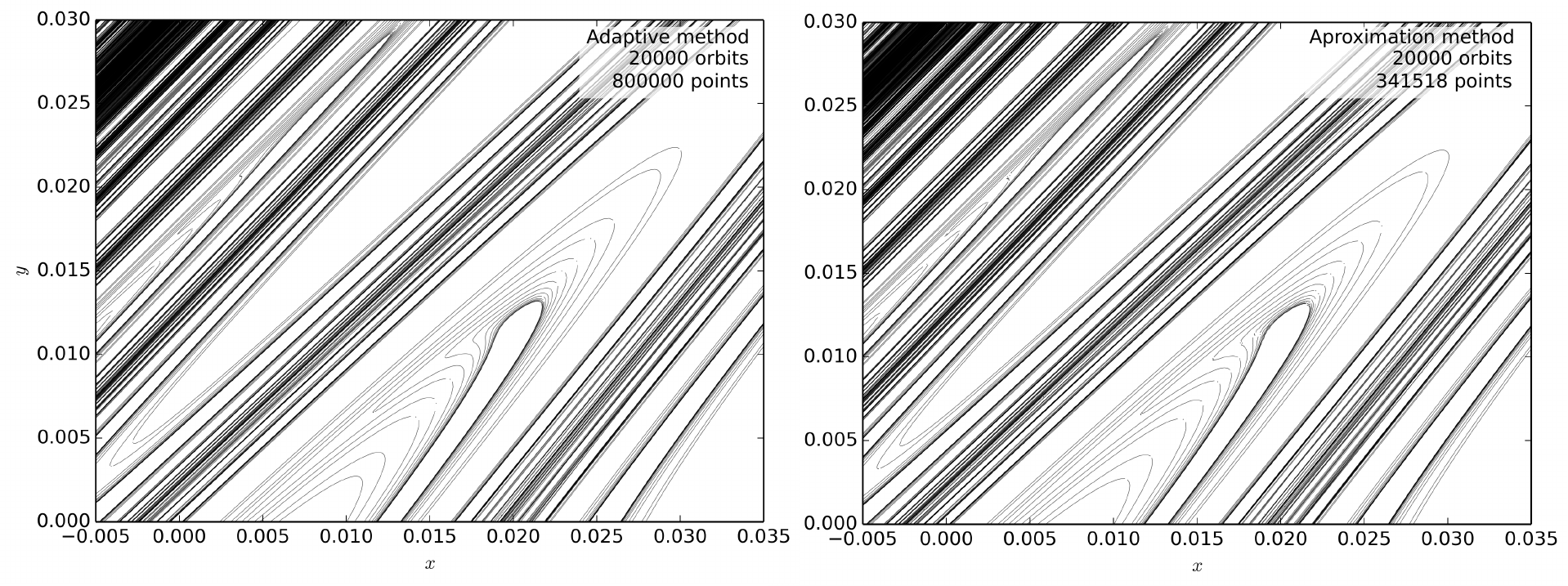}
 \caption{\label{comparison3}Detail of the unstable manifold for the exact adaptive method and the approximated method for $\alpha_{max}=10^\circ,  d_{max}=0.005$ and $d_{tol}=0.05$.}
\end{figure}

\begin{figure}[h]
 \centering
 \includegraphics[width=0.6\textwidth]{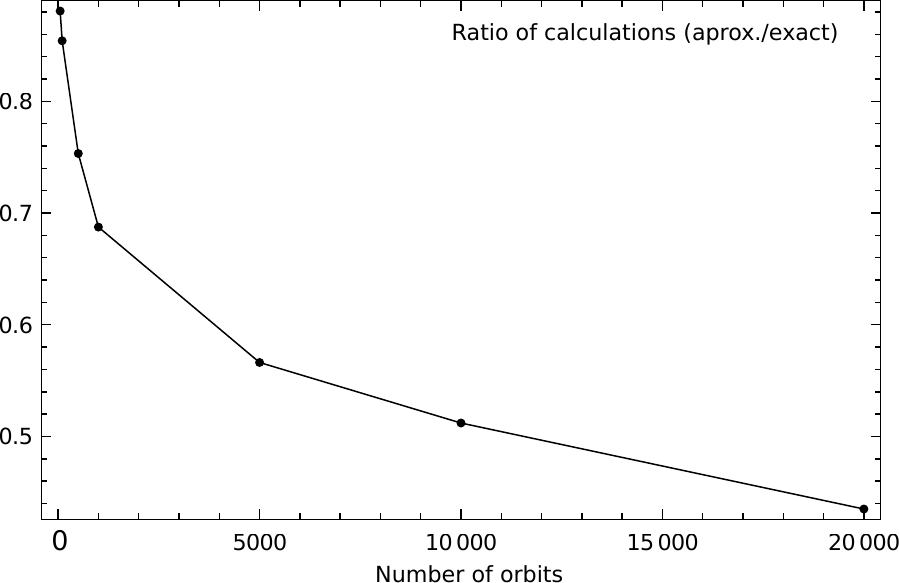}
 \caption{\label{efficiency}Ratio between the number of Poincaré cycles for the approximated method and the number of the exact one. The efficiency of the approximated method grows with the number of orbits.}
\end{figure}

In Fig.~\ref{comparison3} we show the same detail of Fig.~\ref{comparison2} comparing the exact and the approximated adaptive methods for 20000 orbits. Here we obtain the same results with a $57\%$ reduction in the calculations. As we increase the length of the manifold and the number of orbits this method becomes more efficient. In Fig.~\ref{efficiency} we show the ratio between the number of Poincaré cycles required to build manifold with the approximated method and the number of cycles required for the exact calculation.

Another improvement of the approximated method consists in calculating the intermediate point at a previous position in the manifold, for instance, one cycle before the interval where the distance condition is violated. To do this, we can search for the corresponding tags back in the manifold. This will only add an extra point to each new orbit and have a dramatic impact in the long term precision of the approximation. 

In situations where the construction of the Poincaré map is computationally expensive, for instance, when we use an implicit numerical integrator or a predictor-corrector, the approximation of the manifold becomes a more attractive solution, leading to reductions in the computation time of a factor three in regular cases and more when we require longer manifolds. Another important advantage of the approximated method is that it leads to a natural approach to deal with open systems where we know the vector field in a finite region and we lose field lines at some boundary~\cite{simo11}. This kind of situation is relevant in Plasma Physics, where small asymmetries leads to chaotic magnetic flux losses from the plasma edge~\cite{evans02}.

\section{Conclusions}
In the present work we have introduced an efficient algorithm to construct the invariant manifolds of three-dimensional flows possessing unstable periodic orbits. The method results in an ordered high-resolution and optimizable representation of the invariant surfaces linked to a given asymmetric saddle. Additionally, we have introduced a general method to calculate the asymmetric saddle (unstable periodic orbit) from an initial guess, which is useful in cases where the asymmetric vector field has a dominant symmetric component. The presented method is also valid for dissipative systems and conservative systems in non-canonical form. Finally, we have introduced an approximated version of the method to calculate the manifolds. This version does not require full evolution of the orbits from the neighborhood of the magnetic saddle, but introduces the initial conditions near the places where the points along the manifold become sparse, reducing the number of calculations by $N\log N$. In regular situations the calculation efficiency grows by a factor two or three and is better for especially long calculations.

\section{Acknowledgments}
This material is based upon work supported by the U.S. Department of Energy, Office of Science, Office of Fusion Energy Sciences, using the DIII-D National Fusion Facility, a DOE Office of Science user facility, under DE-FC02-04ER54698 and under DE-SC0012706 and DE-FG02-05ER54809. DIII-D data shown in this paper can be obtained in digital format by following the links at https://fusion.gat.com/global/D3D$\underline{~}$DMP.
This research was depeloped with the financial support of the Brazilian scientific agencies CAPES, CNPq and the S\~ao Paulo Research Foundation (FAPESP) under grants 2011/19269-11, 2012/18073-1 and 2014/03899-7.


\section*{References}
\begin{thebibliography}{10}
\expandafter\ifx\csname url\endcsname\relax
  \def\url#1{\texttt{#1}}\fi
\expandafter\ifx\csname urlprefix\endcsname\relax\def\urlprefix{URL }\fi
\expandafter\ifx\csname href\endcsname\relax
  \def\href#1#2{#2} \def\path#1{#1}\fi

\bibitem{goldstein80}
H.~Goldstein, Classical Mechanics, Addison-Wesley, 1980.

\bibitem{zaslavsky07}
G.~Zaslavsky, The Physics of chaos in Hamiltonian systems, Imperial College
  Press, 2007.

\bibitem{guckenheimer83}
J.~Guckenheimer, P.~Holmes, Nonlinear Oscillations, Dynamical Systems and
  Bifurcations of Vector Fields, Springer - Verlag, 1983.

\bibitem{portela07}
J.~Portela, I.~Caldas, R.~Viana, M.~Sanju\'an, Fractal and wada exit basin
  boundaries in tokamaks, Int. J. Bifurcation Chaos 17~(11) (2007) 4067.

\bibitem{biskamp92}
D.~Biskamp, Nonlinear Magnetohydrodynamics, Cambridge University Press, 1992.

\bibitem{kadomtsev92}
B.~Kadomtsev, Tokamak Plasma: A Complex Physical System, IOP Publishing, 1992.

\bibitem{da-silva02}
E.~da~Silva, I.~Caldas, R.~Viana, M.~Sanju\'an, Escape patterns, magnetic
  footprints, and homoclinic tangles due to ergodic magnetic limiters, Phys.
  Plasmas 9~(12) (2002) 4917.

\bibitem{hobson92}
D.~Hobson, An efficient method for computing invariant manifolds of planar
  maps, Journal of Computational Physics 104 (1993) 14.

\bibitem{simo11}
C.~Sim\'o, A.~Vieiro, Dynamics in chaotic zones of area preserving maps: Close
  to separatrix and global instability zones, Physica D 240 (2011) 732.

\bibitem{evans02}
T.~Evans, R.~Moyer, P.~Monat, Modeling of stochastic magneti flux loss from the
  edge of a poloidally diverted tokamak, Phys. Plasmas 9~(12) (2002) 4597.

\end{thebibliography}
\end{document}